\documentclass[aps,prd,amsmath,amssymb,showpacs]{revtex4}
\usepackage{graphicx}
\usepackage{dcolumn}
\usepackage{morefloats}
\usepackage{color}
\usepackage{slashed}
\usepackage{bbm}

\begin{document}
\title{The vertex corrections due to the triangle singularity mechanism in the light axial vector meson couplings to $K^*\bar{K}+c.c.$}

\author{Meng-Chuan Du$^{1,2}$\footnote{{\it E-mail address:} dumc@ihep.ac.cn}, Yin Cheng$^{1,2}$\footnote{{\it E-mail address:} chengyin@ihep.ac.cn},
and Qiang Zhao$^{1,2}$\footnote{{\it E-mail address:} zhaoq@ihep.ac.cn}
}
\affiliation{$^1$ Institute of High Energy Physics,\\
         Chinese Academy of Sciences, Beijing 100049, China}

\affiliation{$^2$ University of Chinese Academy of
Sciences, Beijing 100049, China}

\begin{abstract}

The light axial vector mesons can couple to $K^*\bar{K}+c.c.$ in an $S$ wave at the tree level. Due to the near-threshold $S$-wave interactions the couplings can be affected by the final state interactions. It is of peculiar interest that the pion exchange between $K^*\bar{K}+c.c.$ can go through a triangle diagram which is within the kinematics of the triangle singularity (TS). This mechanism will introduce energy dependence and a $D$-wave amplitude to the vertex couplings. Hence, it should be necessary to investigate the role played by the $K^*\bar{K}+c.c.$  threshold in order to have a better understanding of the light axial vector spectrum.

\end{abstract}

\maketitle
\section{Introduction}

In the constituent quark model, the light axial vector mesons are categorized as the $P$-wave quark and anti-quark systems with $J^{PC}=1^{+(+)}$ and $1^{+(-)}$ while the positive and negative charge conjugate parities indicate the total spins of the quark and anti-quark systems to be either parallel or anti-parallel. Although the light flavor SU(3) nonets for these two axial vectors have been established in experiment~\cite{Zyla:2020zbs}, it seems that our knowledge about these structures is still far from satisfactory.

In the Particle Data Group (PDG)~\cite{Zyla:2020zbs}, the assignment of the non-strange multiplets are $f_1(1285)$, $f_1(1420)$, $a_1(1260)$ for the $1^{++}$ sector and $h_1(1170)$, $h_1(1415)$ and $b_1(1235)$ for the $1^{+-}$ sector. The strange multiplets can mix with each other and form $K_1(1270)$ and $K_1(1400)$ as their mass eigenstates. 
Although there have been a lot of theoretical and experimental studies on these states, there are still some puzzling issues about their nature and decays. In particular, the open $K^*\bar{K}$ threshold phenomenon is relevant. In Ref.~\cite{Longacre:1990uc} a unitary isobar model is constructed to describe the three-body interaction of the $K\bar{K}\pi$ system where $f_1(1420)$ can be described as a $K\bar{K}\pi$ molecule. In Ref.~\cite{Roca:2005nm} with a Weinberg-Tomozawa term for the vector-pseudoscalar scattering, poles corresponding to $f_1(1285)$, $a_1(1260)$, $h_1(1170)$, $h_1(1415)$, $b_1(1235)$ and $K_1(1270)$ are found on the second Rieman sheet of scattering amplitudes, which suggests molecular nature of these states. However, $f_1(1420)$ and $K_1(1400)$ are unlikely to be accommodated by the molecular picture.

A peculiar threshold phenomenon is the triangle singularity (TS) which was first identified by Landau~\cite{Landau:1959fi} in the final state rescatterings with a $t$-channel particle exchange to form a triangle loop. It occurs in such a kinematic region where all the internal particles of the triangle loop can be simultaneously on-shell, and can result in a characteristic logarithmic singularity for the loop transition amplitude. Although this phenomenon was predicted long time ago, its effect was observed until quite recently. In Ref.\cite{Wu:2011yx} the TS effects due to the $K^*\bar{K}+c.c.$ rescattering by exchanging a Kaon was proposed to explain the large isospin violation in $J/\psi\to\gamma\eta(1405/1475)\to\gamma+3\pi$~\cite{BESIII:2012aa}. In the followed-up extensive studies~\cite{Wu:2012pg,Aceti:2012dj,Achasov:2015uua,Du:2019idk,Cheng:2021nal} the detailed analyses including possible mechanisms have provided crucial information for understanding the pseudoscalar states $\eta(1405/1475)$. In Ref.~\cite{Wu:2012pg} it is shown that the angular distribution of the $\pi$ recoiling against $f_0(980)$ in $J/\psi\to\gamma+3\pi$ requires the $f_1(1420)$ contribution, where the open $S$-wave $K^*\bar{K}$ threshold effect via the TS mechanism is important. Besides, the TS also provides a natural explanation to the $a_1(1420)$ observed in $f_0(980)\pi$ final state in $\pi^- p$ scattering~\cite{Liu:2015taa,Du:2021zdg,Ketzer:2015tqa,Aceti:2016yeb} observed by the COMPASS Collaboration~\cite{Adolph:2015pws}. Recent studies of the manifestations of the TS mechanism can be found in Refs.~\cite{Liu:2015taa} and recent reviews of Refs.~\cite{Guo:2017jvc,Guo:2019twa}.

It is proposed by Ref.~\cite{Debastiani:2016xgg} that $f_1(1420)$ is not a genuine state, but a kinematic effect  of the TS in the decay of $f_1(1285)$. In the $\eta\pi\pi$ decay channel, the line shape of the spectrum based on this scenario  is indeed consistent with the WA102 observation~\cite{Barberis:1998by}. However, in the $K\bar{K}\pi$ channel, the calculation of Ref.~\cite{Debastiani:2016xgg} gives a flat and wide enhancement near the mass of $f_1(1420)$, which is inconsistent with the strongly enhanced sharp peaks observed in various measurements~\cite{Armstrong:1984rn,Armstrong:1989hy,Barberis:1997vf,Abdallah:2003gu}. The TS effect followed by the $K^*\bar{K}$ open $S$-wave channel with $J^{PC}=1^{+-}$ is studied in Ref.~\cite{Jing:2019cbw}, where the pole contributions are not included. In Ref.~\cite{Du:2021zdg} a comprehensive study on the light axial vector mesons is present by assuming all these axial states are the genuine quark-model states, but affected by the TS. The couplings between the axial vectors (i.e.  $1^{++}$ and $1^{+-}$ states) and $K^*\bar{K}+c.c.$ are assumed to be energy-independent and symmetric under the SU(3) flavour group. This assumption needs further investigations since the pion exchange between $K^*$ and $\bar{K}$ can renormalize the axial vector meson couplings to the $K^*\bar{K}+c.c.$ channel. Nevertheless, since the exchanged pion can be on-shell, the occurence of the TS may cause highly non-trivial effects. This motivates us to investigate the axial vector meson couplings to $K^*\bar{K}+c.c.$ dressed by the TS mechanism and it should provide further information for our understanding of the nature of these axial vector states.

In Sec.~\ref{sec2}, we will define the axial vector meson couplings to $K^*\bar{K}+c.c.$ and provide the formalism of the transition amplitudes with the triangle loop corrections. In Sec.~\ref{sec3} we will present the numerical results and discuss the impact of the triangle loop corrections on the vertex couplings. Conclusions will be made in the last Section.

\section{Axial vector meson couplings to $K^*\bar{K}+c.c.$ \label{sec2}}

\subsection{Tree-level couplings}

The non-strange light axial vector mesons, $f_1(1420)$ and $h_1(1415)$, can couple to $K^*\bar{K}$ in an $S$ wave, and then decay into $K\bar{K}\pi$. In this work we focus on the these two states since their masses are within the TS kinematic region. We mention in advance that the TS corrections to the lower mass states $f_1(1285)$, $a_1(1260)$, $h_1(1170)$, and $b_1(1235)$ are small and can be neglected. Meanwhile, the $a_1(1420)$ can be regarded as a non-resonance structure caused by the TS.

\begin{figure}[htbp]
\includegraphics[width=5in]{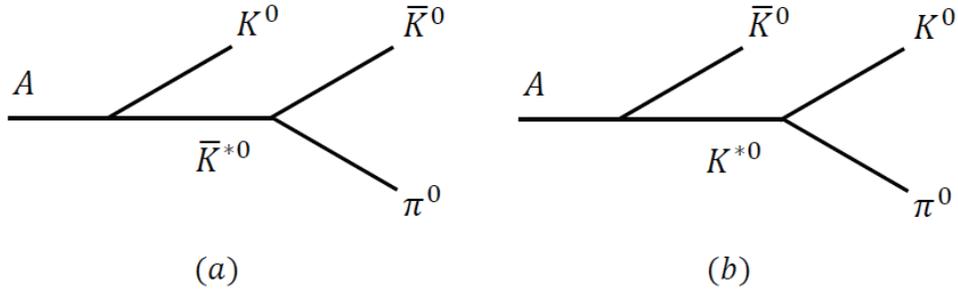}
\caption{Tree-level transitions of $A\to K^0\bar{K}^0\pi^0$ where $A=f_1$ or $h_1$ are implied. }\label{diagram_AtoK0K0barpi0_tree}
\end{figure}

In Fig.~\ref{diagram_AtoK0K0barpi0_tree} the tree-level processes for $A \to K^*\bar{K}+c.c.\to K^0\bar{K}^0\pi$ are illustrated, where $A$ represents the non-strange axial vector states $f_1$ with $J^{PC}=1^{++}$ or $h_1$ with $J^{PC}=1^{+-}$, respectively. A bare interaction between $A$ and $K^*\bar{K}+c.c.$ respecting the SU(3) flavor symmetry is introduced in Fig.~\ref{diagram_AtoK0K0barpi0_tree}, i.e.
\begin{eqnarray}\label{eq_AVPBVP}
L_{f_1VP}&=&g_{f_1VP}\langle f_1^{\mu}[V_{\mu},P] \rangle\ , \nonumber\\
L_{h_1VP}&=&ig_{h_1VP}\langle h_1^{\mu}\{V_{\mu},P\}\rangle \ ,
\end{eqnarray}
where $\langle\dots\rangle$ represents the trace of the matrix inside, and $V$ and $P$ are the vector and pseudoscalar meson fields, respectively; $g_{f_1VP}$ and $g_{h_1VP}$ are the corresponding coupling constants.  
The transition of $K^*\to K\pi$ is described by 
\begin{equation}
L_{VPP}=ig_{VPP}\langle V^{\mu}[\partial_{\mu}P,P]\rangle,\label{eq_VPP}
\end{equation}
with $g_{VPP}=4.52$ calculated by the $\phi\to K\bar{K}$ or $K^*\to K\pi$ decays. The SU(3) multiplets of the vector and pseudoscalar mesons are given respectively as follows:
\begin{eqnarray}
V^{\mu}&\equiv &\left(
\begin{array}{ccc}
  \frac{\rho^0}{\sqrt{2}}+\frac{\omega}{\sqrt{2}} & \rho^+ & K^{*+} \\
  \rho^- & \frac{-\rho^0}{\sqrt{2}}+\frac{\omega}{\sqrt{2}} & K^{*0} \\
  K^{*-} & \bar{K}^{*0} & \phi
\end{array}
\right),
\end{eqnarray}
and 
\begin{eqnarray}
P&\equiv &\left(
\begin{array}{ccc}
  \frac{\pi^0}{\sqrt{2}}+\frac{\cos{\alpha_P}\eta+\sin{\alpha_P}\eta'}{\sqrt{2}} & \pi^+ & K^+ \\
  \pi^- & -\frac{\pi^0}{\sqrt{2}}+\frac{\cos{\alpha_P}\eta+\sin{\alpha_P}\eta'}{\sqrt{2}} & K^0 \\
  K^- & \bar{K}^0 & \sin{\alpha_P}\eta+\cos{\alpha_P}\eta'
\end{array}
\right),
\end{eqnarray}
where the mixing between $\eta$ and $\eta'$ is defined as
\begin{eqnarray}\label{pseudoscalar-mixing}
\left(
  \begin{array}{c}
    \eta \\
    \eta' \\
  \end{array}
\right)
=
\left(\begin{array}{cc}
\cos{\alpha_P} & -\sin{\alpha_P}\\
\sin{\alpha_P} & \cos{\alpha_P}
\end{array}
\right)
\left(
  \begin{array}{c}
    \eta_n \\
    \eta_s \\
  \end{array}
\right) \ ,
\end{eqnarray}
with $\eta_n\equiv (u\bar{u}+d\bar{d})/\sqrt{2}$ and $\eta_s\equiv s\bar{s}$, and $\alpha_P$ is the mixing angle.

The tree-level amplitudes of Fig.~\ref{diagram_AtoK0K0barpi0_tree} (a) and (b) can be obtained:
\begin{eqnarray}
M^{tree}_1&=&g_{tree1}\epsilon_{\mu}\frac{i(-g_{\mu\nu}+\frac{(p_b+p_d)_{\mu}(p_b+p_d)_{\nu}}{s_{bd}})}{s_{bd}-m_{K^*}^2+im_{K^*}\Gamma_{K^*}}i(p_b-p_d)^{\nu}\nonumber\\
&=&-\frac{1}{s_{bd}-m_{K^*}^2+im_{K^*}\Gamma_{K^*}}g_{tree1}\epsilon_{\mu}[(-1+\frac{s_b-s_d}{s_{bd}})p_b^{\mu}+(1+\frac{s_b-s_d}{s_{bd}})p_d^{\mu}],\label{eq_ABtree1}\\
M^{tree}_2&=&g_{tree2}\epsilon_{\mu}\frac{i(-g_{\mu\nu}+\frac{(p_a+p_d)_{\mu}(p_a+p_d)_{\nu}}{s_{ad}})}{s_{ad}-m_{K^*}^2+im_{K^*}\Gamma_{K^*}}i(p_a-p_d)^{\nu},\nonumber\\
&=&-\frac{1}{s_{ad}-m_{K^*}^2+im_{K^*}\Gamma_{K^*}}g_{tree2}\epsilon_{\mu}[(-1+\frac{s_a-s_d}{s_c})p_a^{\mu}+(1+\frac{s_a-s_d}{s_c})p_d^{\mu}],\label{eq_ABtree2}
\end{eqnarray}
where the momenta of $\bar{K}^0$, $K^0$, and $\pi^0$ are labelled by $p_a$, $p_b$ and $p_d$, respectively. The invariant masses squared of the $K^0\pi$ and $\bar{K}^0\pi$ subsystem are denoted by $s_{ad}=(p_a+p_d)^2$ and $s_{bd}=(p_b+p_d)^2$, respectively, and a constant width for $K^*$ meson is adopted, i.e. $\Gamma_{K^*}=50$ MeV. The products of coupling constants in Fig.~\ref{diagram_AtoK0K0barpi0_tree} (a) and (b) are grouped into $g_{tree1}$ and $g_{tree2}$, respectively. For instance, for the initial state $A$ ($A=f_1$ or $h_1$), we define
\begin{eqnarray}
g_{tree1}&\equiv &i\frac{1}{\sqrt{2}} g_{AK^{*0}\bar{K}^0}g_{VPP},\\
g_{tree2}&\equiv &-i\frac{1}{\sqrt{2}}g_{A\bar{K}^{*0}K^0}g_{VPP}.
\end{eqnarray}
Note that for the $C=+1$ state, e.g. $f_1(1420)$, one has $g_{tree1}=g_{tree2}$, but for the $C=-1$ state, e.g. $h_1(1415)$, a sign should be included, i.e. $g_{tree1}=-g_{tree2}$. Since the axial vector states with $C=\pm 1$ can decay into the same $K\bar{K}\pi$ final state, the interference effect can manifest itself at the intersection of the $K^*$ and $\bar{K}^*$ resonance bands in the Dalitz plot.

\subsection{Triangle loop corrections}

\begin{figure}[htbp]
\includegraphics[width=5in]{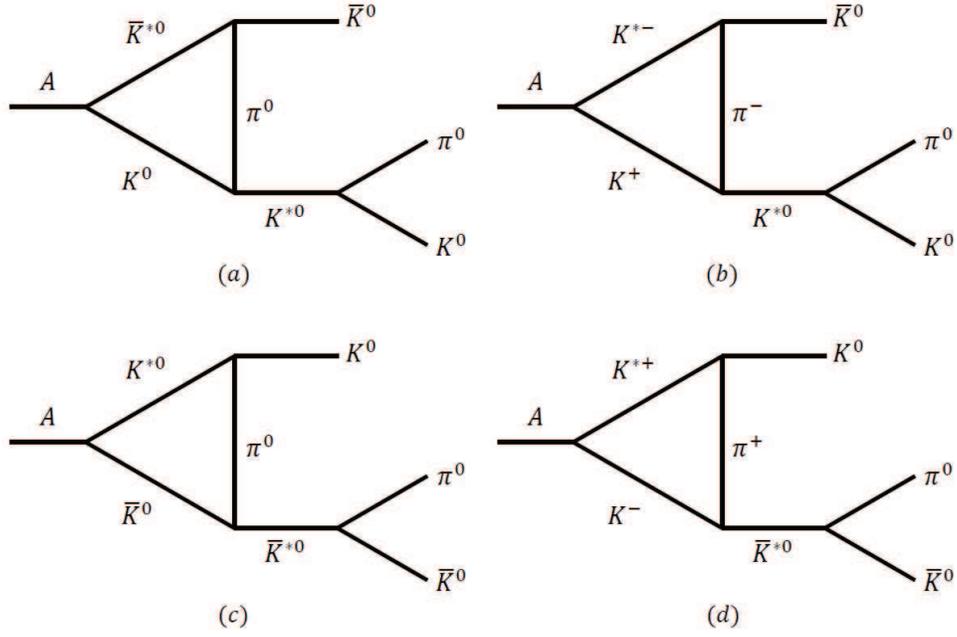}
\caption{Triangle diagrams of $A\to K^0\bar{K}^0\pi^0$ through $\pi$ exchange.}\label{diagram_AtoK0K0barpi0_tri}
\end{figure}

As mentioned earlier, the $S$-wave coupling in Eq.~(\ref{eq_AVPBVP}) is subject to the dressing by the $\pi$ exchange between the intermediate $K^*$ and $\bar{K}$ mesons. As shown by Fig.~\ref{diagram_AtoK0K0barpi0_tri} the triangle diagrams can contribute to the couplings of the axial vector mesons to $K^*\bar{K}+c.c.$ Taking into account that the transition processes in Fig.~\ref{diagram_AtoK0K0barpi0_tri} satisfy the TS condition, it is interesting to find out how important is the triangle loop correction and how it should affect the line shapes in the invariant mass spectra.

With the interactions given in Eqs.~(\ref{eq_AVPBVP}) and (\ref{eq_VPP}), the amplitudes of Fig.~\ref{diagram_AtoK0K0barpi0_tri} read
\begin{eqnarray}
M^{tri}_a&=&\epsilon_{\mu}I_a^{\mu\alpha}\frac{i(-g_{\alpha\beta}+\frac{(p_b+p_d)_{\alpha}(p_b+p_d)_{\beta}}{s_{bd}})}{s_{bd}-m_{K^*}^2+im_{K^*}\Gamma_{K^*}}i(p_b-p_d)^{\beta}g_a,\label{eq_Mtria}\\
M^{tri}_b&=&\epsilon_{\mu}I_b^{\mu\alpha}\frac{i(-g_{\alpha\beta}+\frac{(p_b+p_d)_{\alpha}(p_b+p_d)_{\beta}}{s_{bd}})}{s_{bd}-m_{K^*}^2+im_{K^*}\Gamma_{K^*}}i(p_b-p_d)^{\beta}g_b,\label{eq_Mtrib}\\
M^{tri}_c&=&\epsilon_{\mu}I_c^{\mu\alpha}\frac{i(-g_{\alpha\beta}+\frac{(p_a+p_d)_{\alpha}(p_a+p_d)_{\beta}}{s_{ad}})}{s_{ad}-m_{K^*}^2+im_{K^*}\Gamma_{K^*}}i(p_a-p_d)^{\beta}g_c,\label{eq_Mtric}\\
M^{tri}_d&=&\epsilon_{\mu}I_d^{\mu\alpha}\frac{i(-g_{\alpha\beta}+\frac{(p_a+p_d)_{\alpha}(p_a+p_d)_{\beta}}{s_{ad}})}{s_{ad}-m_{K^*}^2+im_{K^*}\Gamma_{K^*}}i(p_a-p_d)^{\beta}g_d,\label{eq_Mtrid}
\end{eqnarray}
where the subscripts  denote the corresponding processes in Fig.~\ref{diagram_AtoK0K0barpi0_tri}.

In each diagram, the couplings from all vertices are grouped into $g_{a,b,c,d}$, i.e. 
\begin{eqnarray}
&&g_a=\frac{1}{\sqrt{2}}g_b=\frac{i}{2\sqrt{2}}g_{A\bar{K}^{*0}K^0}g_{VPP}^3,\\
&&g_c=\frac{1}{\sqrt{2}}g_d=-\frac{i}{2\sqrt{2}}g_{AK^{*0}\bar{K}^0}g_{VPP}^3.
\end{eqnarray}
For the initial state with $C=+1$, $g_a=g_c$, while for $C=-1$, $g_a=-g_c$. The tensor loop integral is defined by
\begin{eqnarray}
I_{c,d}^{\mu\alpha}&=&\int\frac{d^4q}{(2\pi)^4}\frac{i^3(-g^{\mu\nu}+\frac{q^{\mu}q^{\nu}}{q^2})i(2p_b-q)_{\nu}(-i)(p_c+2p_b-2q)^{\alpha}\mathcal{F}(q^2)}{(q^2-m_1^2)[(q-p_b)^2-m_2^2][(q-p_b-p_c)^2-m_3^2]}\nonumber\\
&=&-i[\Lambda_0(s_0,s_b,s_c)g^{\mu\alpha}+\Lambda_{cb}(s_0,s_b,s_c)p_c^{\mu}p_b^{\alpha}+\Lambda_{bb}(s_0,s_b,s_c)p_b^{\mu}p_b^{\alpha}],\label{eq_Icd}
\end{eqnarray}
and 
\begin{eqnarray}
I_{a,b}^{\mu\alpha}&=&I_{c,d}^{\mu\alpha}|_{p_b\to p_a,p_a\to p_b}\nonumber\\
&=&-i[\Lambda_0(s_0,s_a,s_{bd})g^{\mu\alpha}+\Lambda_{cb}(s_0,s_a,s_{bd})(p_b+p_d)^{\mu}p_a^{\alpha}+\Lambda_{bb}(s_0,s_a,s_{bd})p_a^{\mu}p_a^{\alpha}] \ ,\label{eq_Iab}
\end{eqnarray}
where $\mathcal{F}(q^2)$ is a monopole form factor to regularize the divergence of the loop integrals, i.e.
\begin{equation}
\mathcal{F}(q^2)=\prod_{i=1}^3\frac{m_i^2-\Lambda_i^2}{q_i^2-\Lambda_i^2} \ ,
\end{equation}
with $q_i$s the momenta of internal lines as functions of $q$ and $\Lambda_i\equiv m_i+\beta\Lambda_{QCD}$. In the numerical calculations $\Lambda_{QCD}=250$ MeV and $\beta=2$ are adopted.

In Eqs.~(\ref{eq_Icd}) and (\ref{eq_Iab}), the Lorentz scalars $\Lambda_0,\ \Lambda_{cb}$ and $\Lambda_{bb}$ are parametrized out. There are some typical features of the tensor integral $I^{\mu\alpha}$.
\begin{itemize}

\item $I^{\mu\alpha}$ contracts $\epsilon_{\mu}$ and $\epsilon_{K^*\alpha}^*$. The $g^{\mu\alpha}$ term in Eqs.~(\ref{eq_Icd}) and (\ref{eq_Iab}) leads to $\epsilon\cdot\epsilon_{K^*}^*$, so that it is a leading order correction to the bare tree-level $S$-wave interaction. The bare coupling $g_{AK^*\bar{K}}$ will receive a loop correction from $\Lambda_0$, which will change both the absolute value and the phase of $g_{AK^*\bar{K}}$. The physical $S$-wave $A\to K^*\bar{K}$ coupling $g^{eff}$ becomes 
\begin{eqnarray}
g^{eff}_{AK^{*0}\bar{K}^0}=g_{AK^{*0}\bar{K}^0}+g_a'\Lambda_{0a}+g_b'\Lambda_{0b},\\
g^{eff}_{A\bar{K}^{*0}K^0}=g_{A\bar{K}^{*0}K^0}+g_c'\Lambda_{0c}+g_d'\Lambda_{0d}.
\end{eqnarray}
where the subscripts $a,b,c,d$ are used to distinguish the different internal mass configurations in Fig.~\ref{diagram_AtoK0K0barpi0_tri}. The couplings are 
\begin{eqnarray}
&&g_a'=\frac{1}{\sqrt{2}}g_b'=\frac{i}{2}g_{A\bar{K}^{*0}K^0}g_{VPP}^2,\\
&&g_c'=\frac{1}{\sqrt{2}}g_d'=\frac{i}{2}g_{AK^{*0}\bar{K}^0}g_{VPP}^2.
\end{eqnarray}

\item The other two terms in $I^{\mu\alpha}$ are equivalent to a $D$-wave interaction between the axial vector $A$ and $K^*\bar{K}+c.c.$, and will give rise to different distribution in the Dalitz plot of $K\bar{K}\pi$ final state. 

\item The $g^{\mu\alpha}$ term also receives a $D$-wave contribution to the $K^*\bar{K}$ coupling. As a consequence, the $g^{\mu\alpha}$ term will interfere with the rest two terms, even after the whole phase space integration. 
\end{itemize}

To unambiguously define the $S$ and $D$-wave part of $I^{\mu\alpha}$, one has to separate out the pure $S$-wave amplitude in $\Lambda_0$, which does not interfere with the rest terms when the whole amplitude is squared. This separation is unique. For $I^{\mu\alpha}_{cd}$, the $S$ and $D$-wave parts read,
\begin{eqnarray}
I^{\mu\alpha}_S&=&-i(\Lambda_0-\frac{\sqrt{s}E_c}{1+\frac{3s_c}{|\vec{p}_c|^2}}(\Lambda_{bb}-\Lambda_{cb}))g^{\mu\alpha} \ ,\label{eq_IS}
\end{eqnarray}
and 
\begin{eqnarray}
I^{\mu\alpha}_D&=&-i\frac{\sqrt{s}E_c}{1+\frac{3s_c}{|\vec{p}_c|^2}}(\Lambda_{bb}-\Lambda_{cb})g^{\mu\alpha}-i(\Lambda_{cb}p_c^{\mu}p_b^{\alpha}+\Lambda_{bb}p_b^{\mu}p_b^{\alpha}) \ ,\label{eq_ID}
\end{eqnarray}
where $E_c$ and $\vec{p}_c$ are respectively the energy and 3-momentum of $\bar{K}^*$ in the rest frame of the initial state.
With Eqs.~(\ref{eq_IS}) and (\ref{eq_ID}), we are able to define the pure $S$ and $D$-wave amplitudes $M_S$ and $M_D$, similar to Eqs.~(\ref{eq_Mtria})-(\ref{eq_Mtrid}). It can be immediately seen from Eqs.~(\ref{eq_IS}) and (\ref{eq_ID}) that the new term proportional to $|\vec{p}_c|^2$ is suppressed near threshold, which is indeed a typical behaviour of a $D$-wave coupling.

For the two-body decay $A\to K^*\bar{K}$, it can be proved that the functions $I^{\mu\alpha}_S$ and $I^{\mu\alpha}_D$ do not interfere. For the three-body decay $A\to K^*\bar{K}+c.c.\to K\bar{K}\pi$, there are contributions from two intermediate charge-conjugate channels, i.e. $K^*\bar{K}$ and $\bar{K}^*K$. For every single channel, e.g. $K^*\bar{K}$, the terms of $I^{\mu\alpha}_S$ and $I^{\mu\alpha}_D$ do not interfere after the phase space integration over the invariant mass of $\bar{K}\pi$. However, the term of $I^{\mu\alpha}_S$ of one channel may interfere with $I^{\mu\alpha}_D$ of the other channel, even though the phase space integration is performed.

The full amplitude for $A\to K^*\bar{K}+c.c.\to K^0\bar{K}^0\pi^0$ reads 
\begin{equation}
M^{full}_{A\to K^*\bar{K}+c.c.\to K^0\bar{K}^0\pi^0}=\sum_{i=a,b,c,d}M^{tri}_i+\sum_{i=1,2}M^{tree}_i \ .\label{eq_Mfull}
\end{equation}

\section{Results and discussions\label{sec3}}

In this Section we present the numerical results for $A\to K\bar{K}\pi$ with $A=f_1(1420)$ or $h_1(1415)$. We first examine the TS effects on the coupling constants and then investigate the TS interfering effects on the line shapes.

\begin{figure}[htbp]
\includegraphics[width=2.5in]{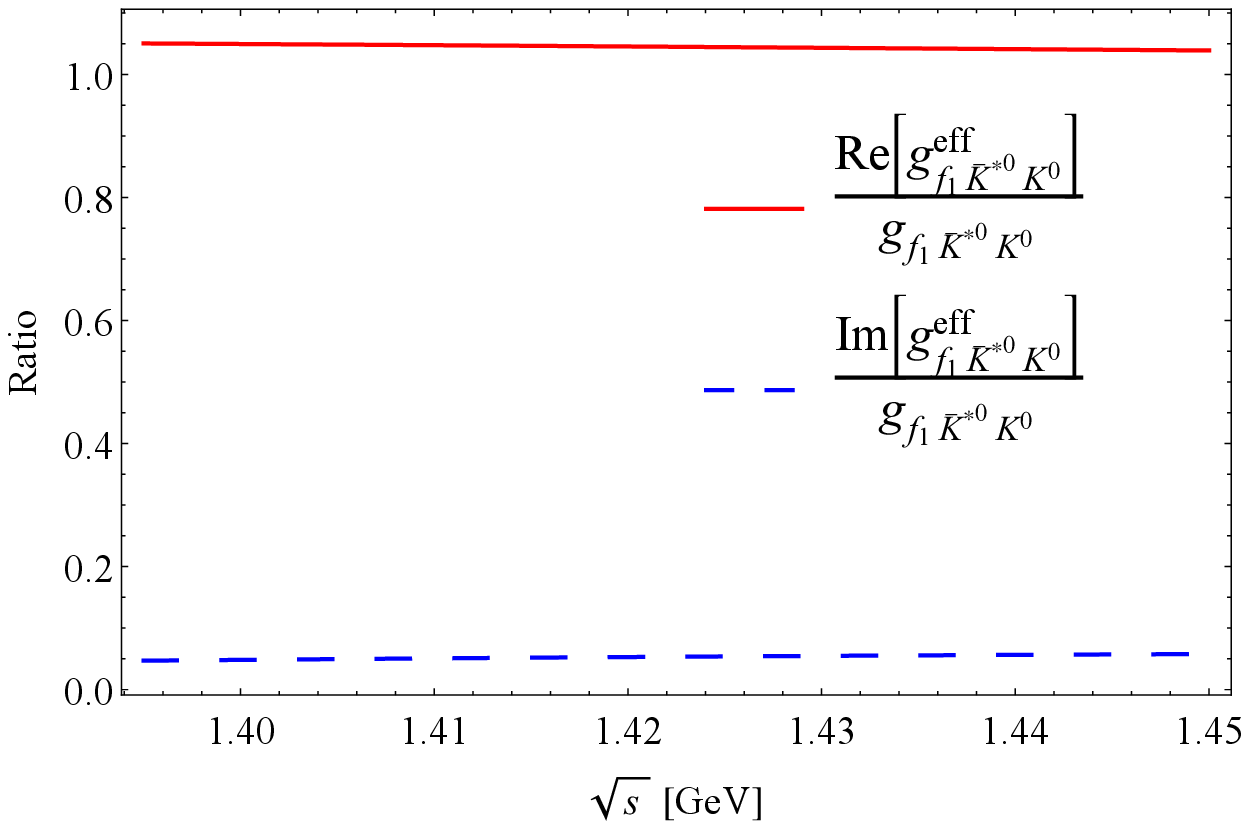}
\includegraphics[width=2.5in]{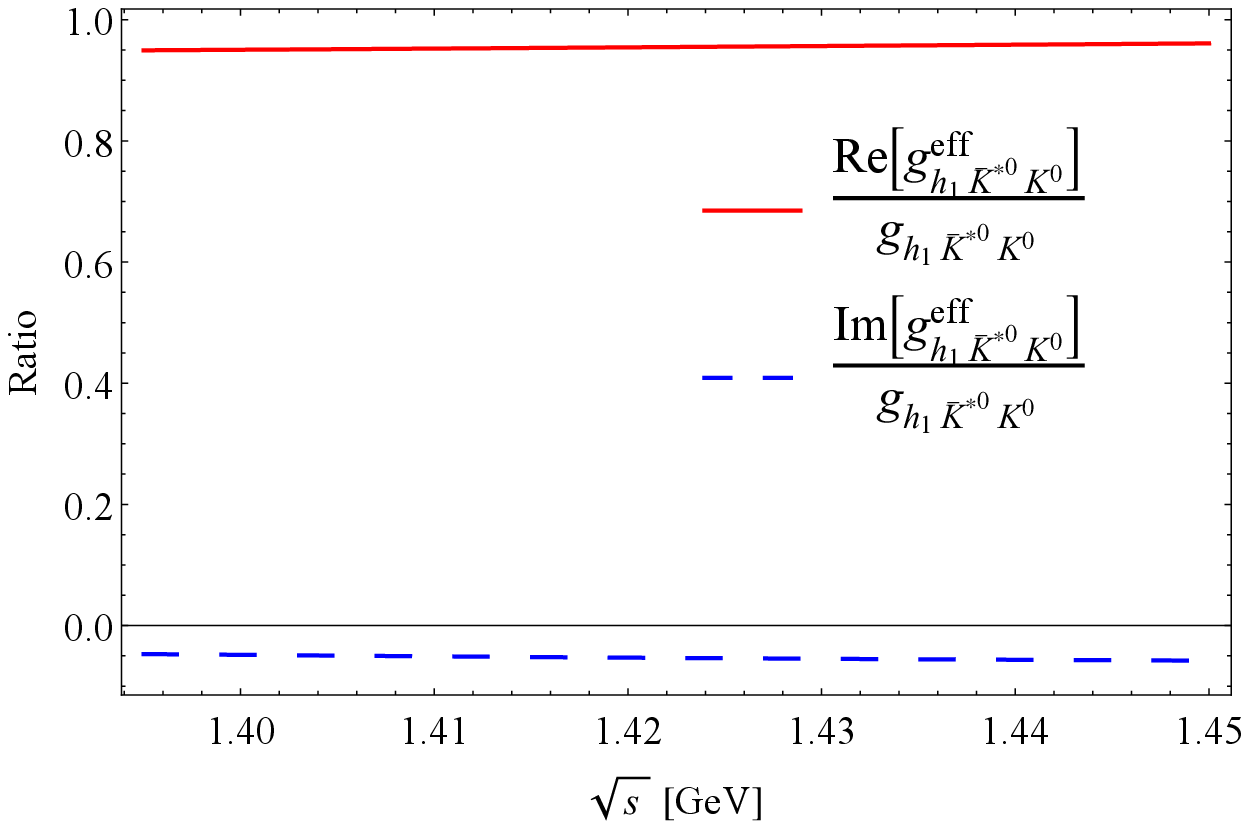}
\caption{The dressed $S$-wave coupling $Re[g^{eff}_{A(B)\bar{K}^{*0}K}]/g_{A(B)\bar{K}^{*0}K}$ (red solid lines) and $Im[g^{eff}_{A(B)\bar{K}^{*0}K}]/g_{A(B)\bar{K}^{*0}K}$ (blue dashed lines) for $f_1(1420)\to \bar{K}^{*0}K$ (left) and $h_1(1415)\to \bar{K}^{*0}K$ (right).}
\label{plot_effcoupling}
\end{figure}

Actually, for the pure $S$-wave tree-level couplings, the triangle diagrams will result in corrections to the $S$-wave coupling constants and introduce a small $D$-wave contributions to the vertex. The bare $S$-wave coupling $g_{AK^*\bar{K}}$ will be shifted to the physical $S$-wave coupling $g^{eff}_{AK^*\bar{K}}$, and because of the loop function,  $g^{eff}_{AK^*\bar{K}}$ is in general a complex number. For the on-shell $K^*$ meson, we evaluate the relative ratios $Re[g^{eff}_{AK^*\bar{K}}]/g_{AK^*\bar{K}}$ and $Im[g^{eff}_{AK^*\bar{K}}]/g_{AK^*\bar{K}}$, for $f_1(1420)$ and $h_1(1415)$, and the results are presented in Fig.~\ref{plot_effcoupling}. It shows that the TS indeed brings non-negligible corrections to the axial vector couplings to $K^*\bar{K}+c.c.$ which are at the order of $5\%$. It also introduces different phases to the physical couplings.

The vertex corrections will result in corrections to the tree-level partial decay width $\Gamma$ for $A\to K\bar{K}\pi$, which is calculated by the tree-level amplitudes (Eqs.~(\ref{eq_ABtree1}) and (\ref{eq_ABtree2})). We define $\delta\Gamma\equiv \Gamma'-\Gamma$, where $\Gamma'$ is the corrected partial width calculated by Eq.~(\ref{eq_Mfull}). Then, the ratio $\delta\Gamma/\Gamma$ will measure the loop correction effects at the physical mass. With the regularization parameter $\beta=2$, we obtain
\begin{eqnarray}
&&\frac{\delta \Gamma(f_1(1420)\to K^*\bar{K}\to K\bar{K}\pi)}{\Gamma(f_1(1420)\to K^*\bar{K}\to K\bar{K}\pi)}=15\%\\
&&\frac{\delta \Gamma(h_1(1415)\to K^*\bar{K}\to K\bar{K}\pi)}{\Gamma(h_1(1415)\to K^*\bar{K}\to K\bar{K}\pi)}=-9.3\%.
\end{eqnarray}
It should be stressed that, the triangle diagrams contain both the $S$ and $D$-wave amplitudes, and $\Gamma'$ is calculated taking into account both contributions. The ratios indicate non-negligible effects arising from the vertex corrections. It shows that inclusions of the vertex correction and the TS effects should be necessary in the partial wave analysis of $A\to K\bar{K}\pi$.

\begin{figure}[htbp]
\includegraphics[width=2.5in]{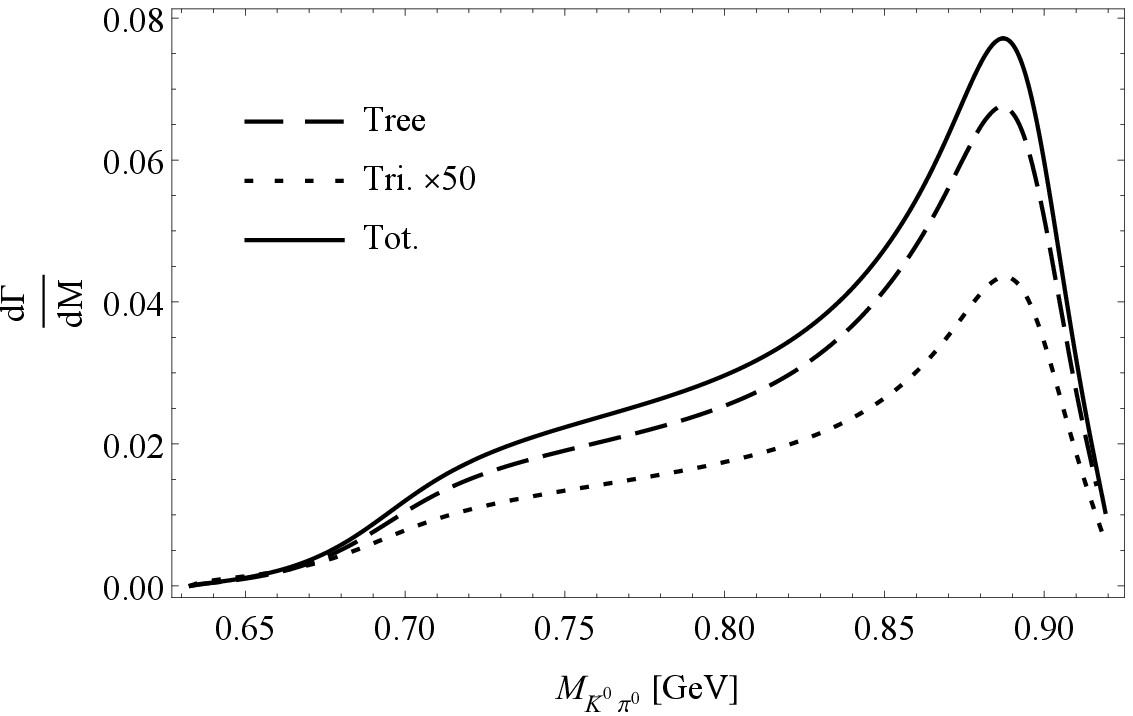}
\includegraphics[width=2.5in]{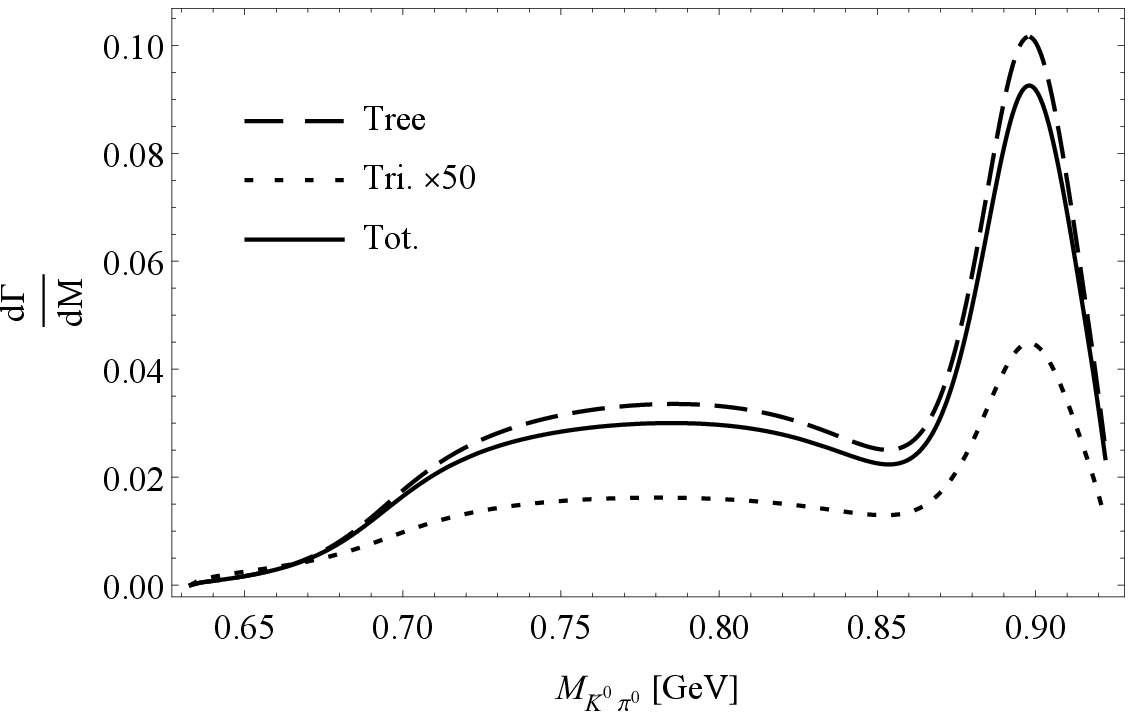}
\caption{The $\bar{K}\pi$ spectra for $f_1(1420)\to K^*\bar{K}+c.c.\to K\bar{K}\pi$ (left) and $h_1(1415)\to K^*\bar{K}+c.c.\to K\bar{K}\pi$ (right). The solid, dashed and dotted lines represent the total, tree and triangle (intensified by 50) contributions, respectively.}\label{spectra_Kpi}
\end{figure}

In Fig.~\ref{spectra_Kpi} the $\bar{K}\pi$ spectra of $f_1(1420)\to K^*\bar{K}+c.c.\to K\bar{K}\pi$ and $h_1(1415)\to K^*\bar{K}+c.c.\to K\bar{K}\pi$ at their physical masses are shown in the left and right panel, respectively. The vertex corrections and triangle loop contributions produce non-trivial spectra. The tree diagrams turn out to be dominant where the $\bar{K}^*$ peaks can be identified. Apart from the $\bar{K}^*$ peaks in both cases, the lower broad bumps are the kinematic reflections from the $K^*\to K\pi$ channel which recoils $\bar{K}$ at the $K^*$ mass. It shows that the correction to the partial width is at the order of $10\%$.

Note that although we calculate the vertex coupling corrections to the axial vector meson couplings to $K^*\bar{K}+c.c.$, we actually show the spectra of the three-body decay channel $K\bar{K}\pi$ via the intermediate $K^*\bar{K}+c.c.$ This is due to the consideration that the interference of the TS may cause changes to the $K^*$ lineshape. 

We also mention that for the final state of $K\bar{K}\pi$ other processes, such as $a_0\pi$ and $\kappa\bar{K}+c.c.$, may contribute in addition to the intermediate $K^*\bar{K}+c.c.$ transition. It means a combined analysis should be necessary in the future with the available data.

\begin{figure}[htbp]
\includegraphics[width=2.5in]{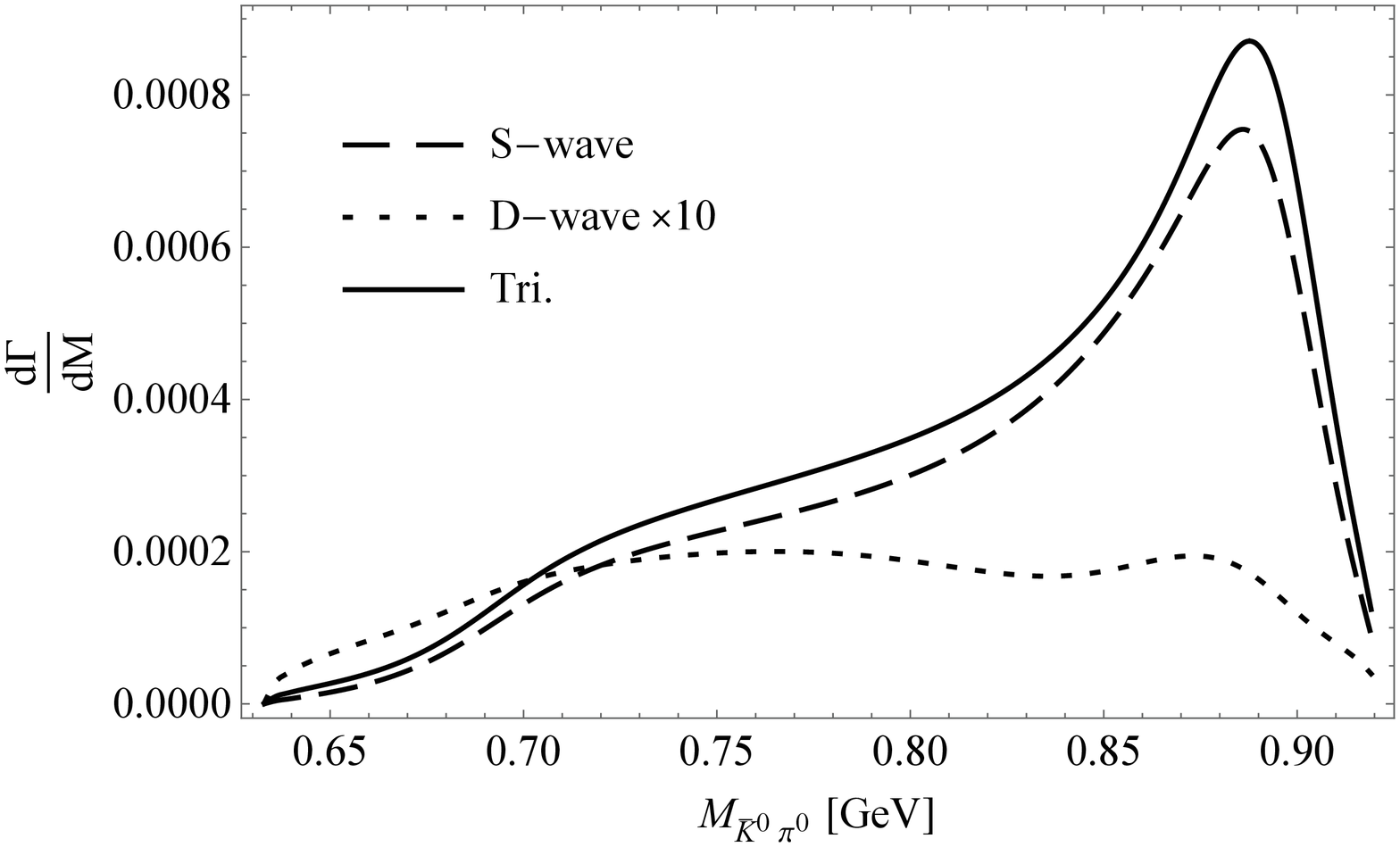}
\includegraphics[width=2.5in]{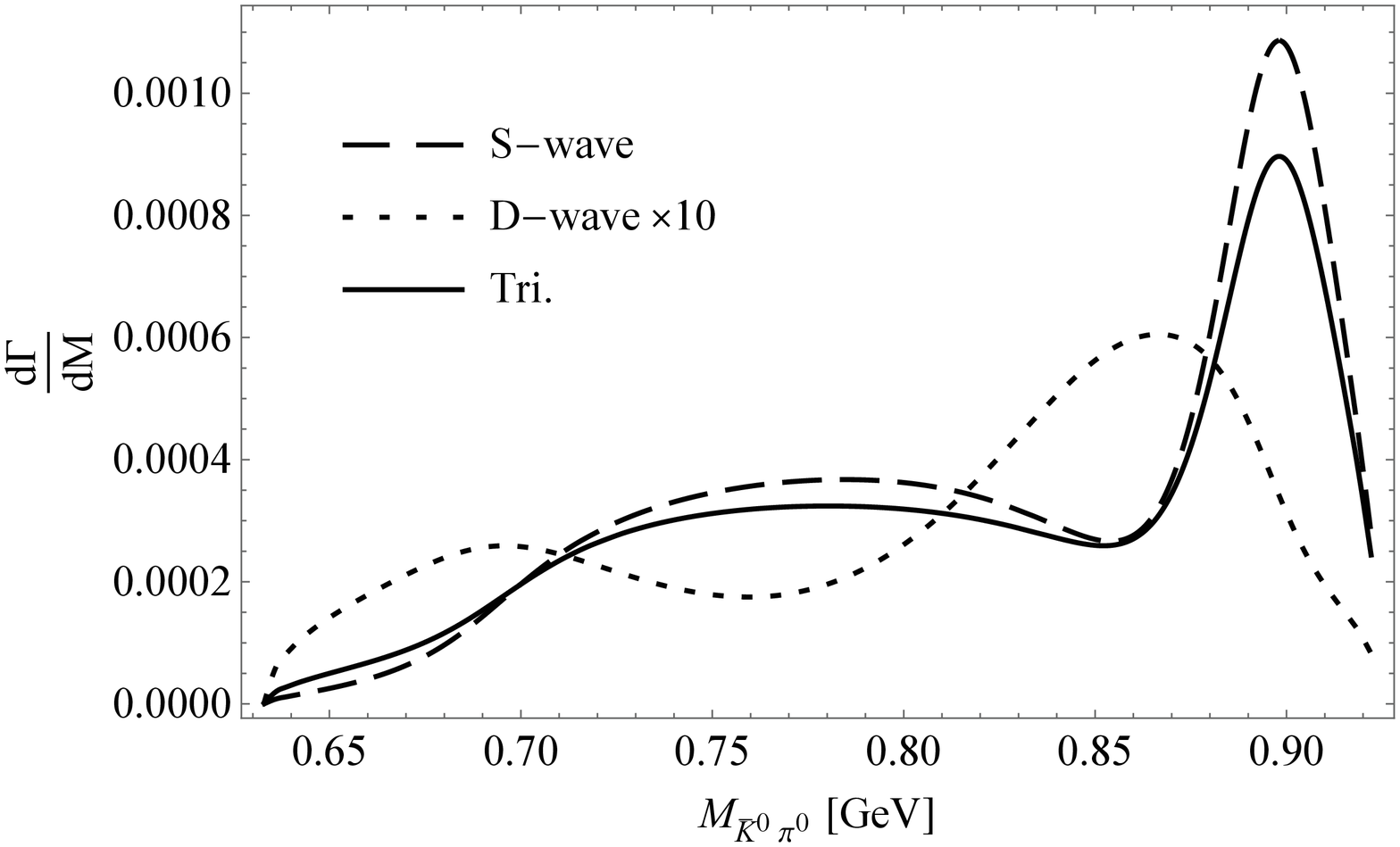}
\caption{The $\bar{K}\pi$ spectra for $f_1(1420)\to K^*\bar{K}+c.c.\to K\bar{K}\pi$ (left) and $h_1(1415)\to K^*\bar{K}+c.c.\to K\bar{K}\pi$ (right) from the triangle loops. The solid lines are full triangle amplitudes. The dashed lines denote the pure $S$-wave contributions from the loop amplitudes, while the dotted lines are the pure $D$-wave contributions multiplied by a factor of 10.}\label{spectra_Kpi_tri}
\end{figure}

Qualitatively, the correction affects the line shape of the two-body spectra rather weakly, but should not be neglected. Nevertheless, the vertex corrections will introduce a $D$-wave amplitude in $I^{\mu\alpha}$.
To quantify the contribution from the $D$-wave part in $I^{\mu\nu}$, we illustrate the individual contributions from the pure $S$-wave (dashed line) and pure $D$-wave (dotted lines) parts in Fig.~\ref{spectra_Kpi_tri}. In both cases for the $f_1(1420)$ and $h_1(1415)$ decays, the $D$-wave amplitude is significantly smaller than the tree-level $S$-wave one, which suggests that the loop corrections will not alter the line shapes of $K\pi$ spectra in Fig.~\ref{spectra_Kpi}. This also means that the vertex corrections do not cause significant changes to the line shapes and a leading-order calculation based on the tree-level $S$-wave couplings are reasonable in the description of the $K\pi$ (or $\bar{K}\pi$) line shapes from the intermediate $K^*\bar{K}+c.c.$ rescatterings~\cite{Du:2021zdg}.

\begin{figure}[htbp]
\includegraphics[width=2.5in]{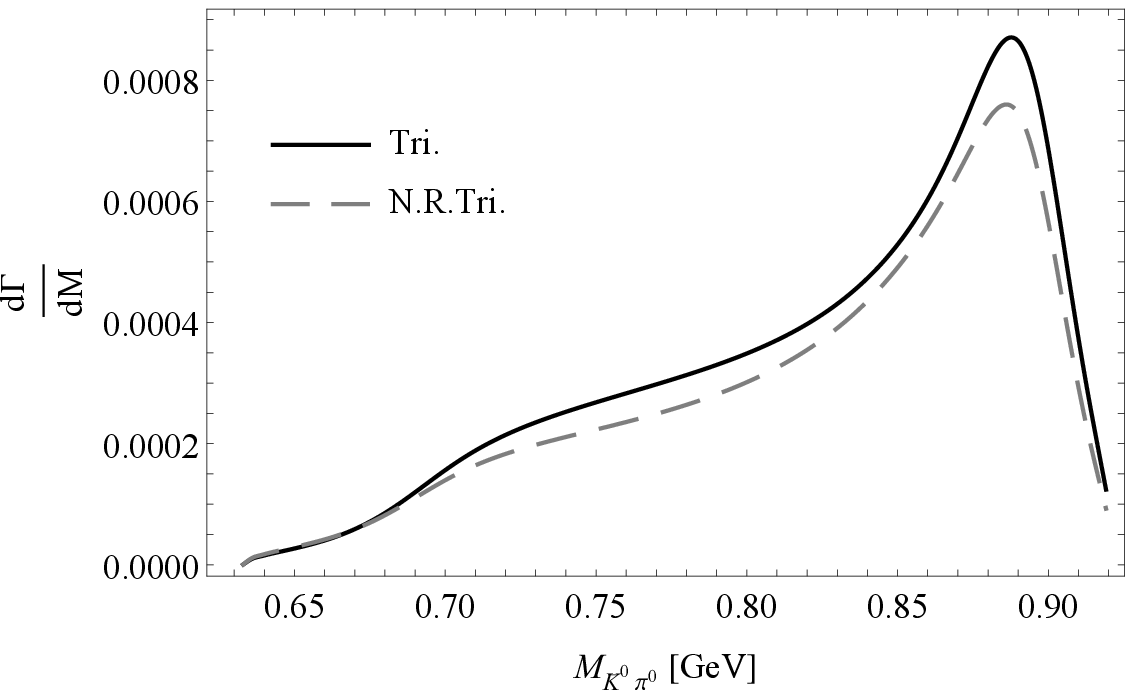}
\includegraphics[width=2.5in]{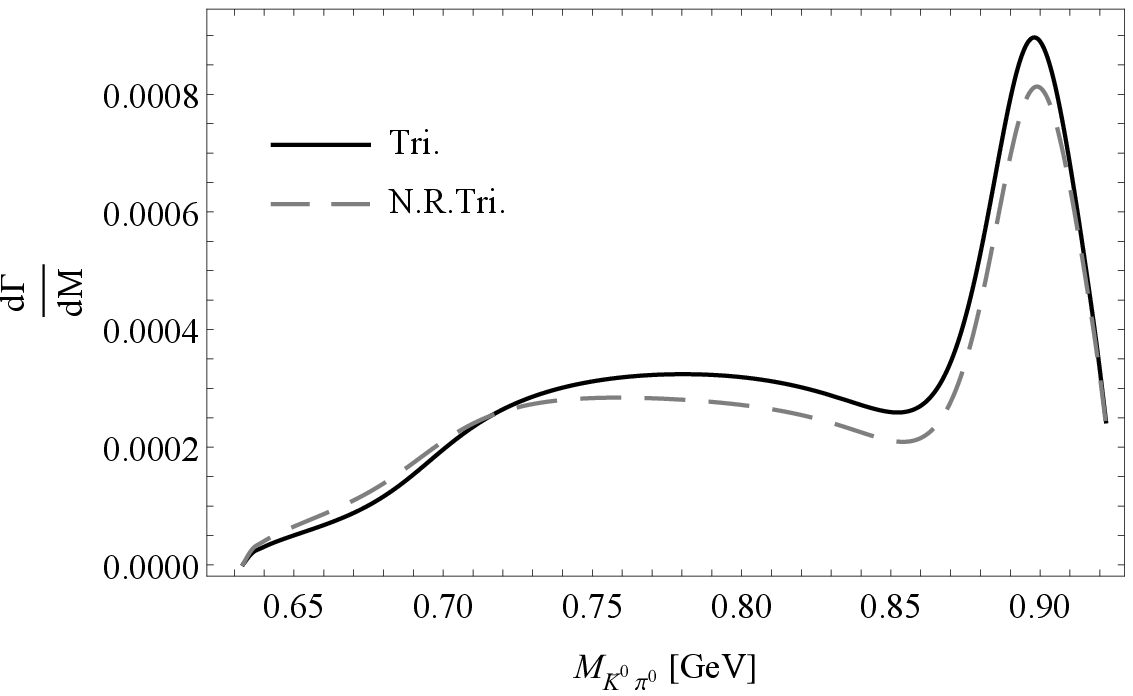}
\caption{The $\bar{K}\pi$ spectra for $f_1(1420)\to K^*\bar{K}+c.c.\to K\bar{K}\pi$ (left) and $h_1(1415)\to K^*\bar{K}+c.c.\to K\bar{K}\pi$ (right) from triangle loops. The solid and dashed lines are calculation results based on the relativistic and non-relativistic formalism, respectively.}\label{spectra_Kpi_tri_NR}
\end{figure}

The dominance of the TS mechanism in the vertex corrections suggests that the main contributions of the triangle loops should come from the kinematic region where these internal particles are nearly on-shell. It allows one to make a non-relativistic expansion to the numerator of Eq.~(\ref{eq_Icd}) by the substitution of $-g^{\mu\nu}+\frac{q^{\mu}q^{\nu}}{q^2}\to\delta^{ij}=-g^{\mu\nu}+g^{0\mu}g^{0\nu}$. 
In this way, the numerator of the integrand in Eq.~(\ref{eq_Icd}) becomes
\begin{eqnarray}
num.\to 4p_b^{\mu}p_b^{\alpha}-4p_b^{\mu}q^{\alpha}-2q^{\mu}p_b^{\alpha}+2q^{\mu}q^{\alpha} , \label{eq_nrnum}
\end{eqnarray}
where the only term leading to the 3-point scalar loop integral $C_0$ is from $4p_b^{\mu}p_b^{\alpha}$. However, one notices that this is a $D$-wave term and proportional to $|\vec{p}_c|^2$. Thus, the amplitude is actually suppressed in the kinematic region near threshold. This explains why the TS effects do not produce significant vertex corrections to the coupling in both $S$ and $D$ waves.  In Fig.~\ref{spectra_Kpi_tri_NR} the results for the $\bar{K}\pi$ invariant mass spectra with the relativistic and non-relativistic amplitudes of the triangle diagrams are compared. The difference indicates the effects of the non-relativistic expansion. One sees that these two results agree to each other quite well.

It is interesting to mention that the triangle loops considered here, namely $K^*\bar{K}(\pi)+c.c.$, are relatively suppressed in comparison with the ones involving $K^*\bar{K} (K)+c.c.$ such as $h_1(1415)\to\phi\pi$~\cite{Du:2021zdg}. For the latter cases, the TS contributions are relatively enhanced by the larger phase spaces.

\section{Conclusion}

We investigate the triangle loop corrections to the axial vector couplings to $K^*\bar{K}+c.c.$ where the TS mechanism can contribute. Due to the triangle loop transitions the tree-level $S$-wave couplings for $f_1(1420)$ and $h_1(1415)$ to   $K^*\bar{K}+c.c.$ will be corrected by about $5\%$, and a $D$-wave coupling can arise from the initial $S$-wave couplings. Although these corrects are rather small, their interferences with the tree-level amplitudes can still produce  some effects on the invariant mass spectra in the final states. In particular, the presence of the TS mechanism within the near-threshold region will change the energy-dependent behaviour of the initial state. It suggests that a proper consideration of the vertex corrections as well as the TS mechanism would be necessary for the study of these axial vector mesons in the future partial wave analysis e.g. in charmonium radiative or hadronic decays at BESIII.

\begin{acknowledgments}
This work is supported, in part, by the National Natural Science Foundation of China (Grant No. 11521505), the DFG and NSFC funds to the Sino-German CRC 110 ``Symmetries and the Emergence of Structure in QCD'' (NSFC Grant No. 12070131001, DFG Project-ID 196253076), National Key Basic Research Program of China under Contract No. 2020YFA0406300, and Strategic Priority Research Program of Chinese Academy of Sciences (Grant No. XDB34030302).
\end{acknowledgments}

\section*{Appendix}
\begin{appendix}

We provide the detailed expressions for the transition amplitudes of the triangle loops here. After contraction, the Lorentz structures of the triangle amplitudes take the following forms: 
\begin{eqnarray}
M^{tri}_{c,d}&=&-\frac{ig_{c,d}}{s_{ad}-m_{K^*}^2+im_{K^*}\Gamma_{K^*}}\epsilon_{\mu}[\chi_a(s_0,s_a,s_b,s_c,s_{ab})p_a^{\mu}+\chi_b(s_0,s_a,s_b,s_c,s_{ab})p_b^{\mu}+\chi_d(s_0,s_a,s_b,s_c,s_{ab})p_d^{\mu}],\label{eq_Mtricdfinal}\\
M^{tri}_{a,b}&=&-\frac{ig_{a,b}}{s_{bd}-m_{K^*}^2+im_{K^*}\Gamma_{K^*}}\epsilon_{\mu}[\chi_b(s_0,s_b,s_a,s_{bd},s_{ab})p_a^{\mu}+\chi_a(s_0,s_b,s_a,s_{bd},s_{ab})p_b^{\mu}+\chi_d(s_0,s_b,s_a,s_{bd},s_{ab})p_d^{\mu}].\label{eq_Mtriabfinal}
\end{eqnarray}

The scalar functions $\chi_r$ ($r=a, \ b, \ d$) are linear combinations of $\{\Lambda_0,\Lambda_{cb},\Lambda_{bb}\}$, i.e.
\begin{eqnarray}
&&\chi_a(s_0,s_a,s_b,s_c,s_{ab})\nonumber\\
&=&(-1+\frac{s_a-s_d}{s_c})\Lambda_0(s_0,s_b,s_c)+\frac{1}{2s_c}[s_0(s_c+s_a-s_d)+s_b(s_c-s_a+s_d)+s_c(s_a+s_d-s_c-2s_{ab})]\Lambda_{cb}(s_0,s_b,s_c),\nonumber\\
&&\chi_b(s_0,s_a,s_b,s_c,s_{ab})\nonumber\\
&=&\frac{1}{2s_c}[s_0(s_c+s_a-s_d)+s_b(s_c-s_a+s_d)+s_c(s_a+s_d-s_c-s_{ab})]\Lambda_{bb}(s_0,s_b,s_c),\nonumber\\
&&\chi_d(s_0,s_a,s_b,s_c,s_{ab})\nonumber\\
&=&(1+\frac{s_a-s_d}{s_c})\Lambda_0(s_0,s_b,s_c)+\frac{1}{2s_c}[s_0(s_c+s_a-s_d)+s_b(s_c-s_a+s_d)+s_c(s_a+s_d-s_c-s_{ab})]\Lambda_{cb}(s_0,s_b,s_c).
\end{eqnarray}

It can be immediately verified that (by setting $\Lambda_{bb}$ and $\Lambda_{cb}$ to be zero), when there is only the $g^{\mu\alpha}$ term in $I^{\mu\alpha}$ (or when other $D$-wave terms are negligible), the Lorentz structures of Eqs.~(\ref{eq_Mtricdfinal}) and (\ref{eq_Mtriabfinal}) will recover the tree-level forms of Eqs.~(\ref{eq_ABtree1}) and (\ref{eq_ABtree2}). In such a case, the triangle amplitude will have only the $S$-wave contributions and will not change the distribution pattern of events in the Dalitz plot.

The coefficients $\Lambda_0, \Lambda_{bb}$ and $\Lambda_{bc}$ are expressed in terms of standard functions of LoopTools, i.e. $C_i$, $C_{ij}$ and $C_{ijk}$. Their expressions are, 
\begin{eqnarray}
&&\Lambda_0=\frac{i}{8\pi^2m_1^2}[2s_b \delta C_{001}+(s+s_b-s_c)\delta C_{002}],\\
&&\Lambda_{cb}=\frac{i}{8\pi^2m_1^2}[2\delta C_{002}+2s_b\delta C_{112}+(s+3s_b-s_c)\delta C_{122}+(s+s_b-s_c)\delta C_{222}+2s_b\delta C_{12}+(s+s_b-s_c)\delta C_{22}],\\
&&\Lambda_{bb}=-\frac{i}{4\pi^2}(C_0+C_1+C_2)\nonumber\\
&&+\frac{i}{8\pi^2m_1^2}[2\delta C_{00}+2s_b\delta C_{11}+(s+3s_b-s_c)\delta C_{12}+(s+s_b-s_c)\delta C_{22}+4\delta C_{001}+4\delta C_{002}+2s_b \delta C_{111}+(s+5s_b-s_c)\delta C_{112}\nonumber\\
&&+(2s+4s_b-2s_c)\delta C_{122}+(s+s_b-s_c)\delta C_{222}],
\end{eqnarray}
where the $\delta C_{ij}, \delta C_{ijk}$ are defined by
\begin{equation}
\delta C_{ij(ijk)}=C_{ij(ijk)}-C_{ij(ijk)}|_{m_1\to 0}.
\end{equation}

\end{appendix}

\bibliographystyle{unsrt}

\end{document}